\documentclass[a4paper]{jpconf}
\usepackage{iopams}

\begin{document}

\title{To make a nanomechanical Schr\"{o}dinger-cat mew}

\author{Tam\'as Geszti}
\address{Department of Physics of Complex Systems, E\"otv\"os University, Budapest, Hungary}

\ead{geszti@elte.hu}

\begin{abstract}
By an explicite calculation of Michelson interferometric output intensities in the optomechanical scheme 
proposed by Marshall {\it et al.} \cite{marshall}, an oscillatory factor is obtained that may go down to zero 
just at the time a visibility revival ought to be observed. Including a properly tuned phase shifter offers a simple 
amendment to the situation. By using a Pockels phase shifter with fast time-dependent modulation in one arm, 
one may obtain further possibilities to enrich the quantum state preparation and reconstruction abilities of the 
original scheme, thereby improving the chances to reliably detect genuine quantum behaviour of a 
nanomechanical oscillator.
\end{abstract}

Detecting genuine quantum effects in nanomechanical systems is a highly desirable scope, expected to be
reached in the near future. The major challenges are: efficient cooling close to the ground state, 
sufficiently strong coupling to a well-identified quantum system, reliable preparation and identification of 
non-classical states of the nanomechanical component. Most of the experimental studies are done on mechanical 
oscillators, considered harmonic for the small displacements characteristic for the quantum domain. For those 
systems ground-state cooling means $k_B T < \hbar\omega$ which is relatively easy to fulfil for hard 
(high-frequency) oscillators \cite{grsc}. For them, however, preparing and analyzing quantum states 
is overly demanding \cite{preprec}. The efforts to reconcile these conflicting requirements have seen rapid and 
most competitive advance with the participation of a number of experimental groups. A still more ambitious,
so far elusive prospect of related studies would be to detect deviations from standard quantum mechanics, 
predicted by several theoretical studies \cite{nonstand}.

An original and promising project, using a soft (low-frequency) nanomechanical oscillator, coupled by a 
mirror to one of the paths of a Michelson interferometer, and using Fabry-Perot cavities 
(on both paths, to preserve interference) to strengthen the optomechanical coupling, has been started by
Marshall {\it et al.} \cite{marshall}, and following extensive development of an optical feedback cooling 
technique \cite{kleckner}, is now approaching the level of seeing quantum effects. 

According to the argument of Ref. \cite{marshall}, for a single-photon source of frequency $\omega_c$, and 
a vibrating mirror of effective mass $M$ and vibration frequency $\omega_m$, acting as one of the mirrors 
delimiting a Fabry-Perot resonator cavity of length $L$, the dimensionless coupling constant is 
$\kappa=(\omega_c/\omega_m)(\sqrt{(\hbar/2M\omega_m)}/L)$. With the vibrating-mirror cavity on 
interferometer arm $A$, and the cavity with two rigid mirrors on arm $B$, starting with a mirror in its 
quantum mechanical ground state $|0\rangle_m$, coupling during time $t$ results in an entangled 
photon-mirror state 
\begin{equation}\label{mirrors}
|\Psi(t)\rangle=\frac{e^{-i\omega_ct}}{\sqrt{2}}
  \left(e^{i\varphi(t)} |A\rangle|\alpha(t)\rangle_m+|B\rangle|0\rangle_m\right)
\end{equation}
where $|A\rangle=|1\rangle_A|0\rangle_B$ and $|B\rangle=|0\rangle_A|1\rangle_B$ denote one-photon 
states with the photon in arm $A$ and in arm $B$ resp., whereas, $|\alpha(t)\rangle_m$ and $|0\rangle_m$ 
are two coherent states of respective complex amplitudes $\alpha(t)$ and $0$ of the vibrating mirror, appearing 
as a pair of Schr\"odinger cat states entangled to the respective orthogonal photon states  $|A\rangle$ 
and $|B\rangle$. 
Finally,
\begin{equation}\label{phases}
\varphi(t)=\kappa^2(\omega_mt-sin\omega_mt);~~~\alpha(t)=\kappa(1-e^{-i\omega_mt}).
\end{equation}

The density matrix  $\hat R(t)=|\Psi(t)\rangle\langle\Psi(t)|$ corresponding to the pure state $|\Psi\rangle$
can be expanded on the photon state basis $|A\rangle=\left(\begin{array}{c}1\\0 \end{array}\right)$,  
$|B\rangle=\left(\begin{array}{c}0\\1 \end{array}\right)$ in the form of a two-by-two matrix
\begin{equation}\label{matrix}
\hat R(t)=\left(\begin{array}{cc}
\hat\rho_{AA}(t)&\hat\rho_{AB}(t)\\\hat\rho_{BA}(t)&\hat\rho_{BB}(t)
\end{array}\right),
\end{equation}
where the matrix elements are operators acting on variables of the vibrating mirror. It has been pointed out  
\cite{marshall,adler} that the crucial quantity characterizing the interference pattern to be observed is the 
nondiagonal element on that basis, $\hat\rho_{AB}(t)=\langle A |\Psi(t)\rangle\langle\Psi(t)|B\rangle$, 
which carries extra information about the motion of the vibrating mirror, missing from the more usual reduced 
density matrix $\hat\rho_m(t)=\hat\rho_{AA}(t)+\hat\rho_{BB}(t)$, obtained by tracing out 
$\hat R(t)$ over photon variables.  

Interference visibility, as expressed by  $\hat\rho_{AB}(t)$, carries a signature of the motion of the vibrating 
mirror, but the usefulness of that signature is seriously limited by interactions of the mirror with the environment.  
Neglecting direct environmental effects on the photon subsystem, the above expansion remains valid, allowing 
decoherence and friction of the vibrating mirror to be included in the dynamics of $\hat\rho_{AB}(t)$ and 
analyzed by standard tools  \cite{adler,bdg}. 

Returning for a moment to the decoherence-free case covered by Eq. (\ref{mirrors}),
let us quote the result of Ref. \cite{marshall}:
\begin{equation}\label{old} 
\Tr_m\hat\rho_{AB}(t)=\frac{e^{i\varphi(t)}}{2}\Tr_m \left[|\alpha(t)\rangle_{m~m}\langle 0|\right]
                    = \frac12 e^{i\varphi(t) - |\alpha(t)|^2/2}, 
\end{equation}
where $\Tr_m$ means trace over mirror variables. From here one concludes that the crucial quantity to keep 
under control is $|\alpha(t)|$, and - using  Eq. (\ref{phases}) - the observation should be carried out at 
multiples of time $t=2\pi/\omega_m$, where $|\alpha(t)|$ returns to $0$ and the interference signal can be 
large enough to be observed, under the extremely stringent condition that decoherence could be suppressed 
by sufficient cooling and isolation from mechanical supports.

That treatment catches the essential point of the phenomenon but fails to give a correct description of the 
underlying oscillatory behaviour, which is relevant for a reliable identification of superposition states of the mirror. 
To show that, we proceed by noticing that Eq. (\ref{mirrors}) is not the final state observed by the photon 
detectors, since the photons pass once more through the beam splitter, before reaching output ports $C$
 (facing $A$) and $D$ (facing $B$). On the time scale of whatever happens to the vibrating mirror, this is a 
fast and coherence-preserving unitary operation on the photon subsystem, transforming the intermediate state 
(\ref{mirrors}) into the final one
\begin{equation}\label{mirrorcat}
\frac12 e^{-i\omega_ct} \left[ 
  |C\rangle\left(|0\rangle_m + i e^{i\varphi(t)}|\alpha(t)\rangle_m\right)   
+ i\,|D\rangle \left( |0\rangle_m - i e^{i\varphi(t)}|\alpha(t)\rangle_m  \right) \right]
\end{equation}
with $|C\rangle=|1\rangle_C|0\rangle_D=(|A\rangle+i|B\rangle)/\sqrt2$ and $|D\rangle
=|0\rangle_C|1\rangle_D=(i|A\rangle+|B\rangle)/\sqrt2$. 
 Carrying out the above unitary transformation on the full density matrix, now we can trace over mirror
 (and eventually, for a full analysis, environmental) variables to obtain directly the density matrix of the 
photons to be detected, in the form
\begin{equation}\label{rhophoton}
\mathbf{R}^{phot}=\left(\begin{array}{cc}
\Tr_m\hat\rho_{CC}&\Tr_m\hat\rho_{CĐ}\\\Tr_m\hat\rho_{DC}&\Tr_m\hat\rho_{DD},
\end{array}\right)
\end{equation}
with the diagonal elements  
\begin{equation}\label{diag}
\left.\begin{array}{c}I_C\\I_D\end{array}\right\}
=\left.\begin{array}{c}\Tr_m\hat\rho_{CC}\\\Tr_m\hat\rho_{DD}\end{array}\right\}
=\frac12 \mp {\rm Im} \left(\Tr_m\hat\rho_{AB}\right)
\end{equation}
describing the output intensities on the two respective ports, and the off-diagonal elements
\begin{equation}\label{offdiag}
\Tr_m\hat\rho_{CD}=\left[\Tr_m\hat\rho_{DC}\right]^*
=\frac12 {\rm Re} \left(\Tr_m\hat\rho_{AB}\right) + i \left(\frac12 - \Tr_m \hat\rho_{AA}\right)
\end{equation}
describing cross-correlations between the two detectors.

Of particular interest are the output intensities, and we see from Eq. (\ref{diag}) that, instead of the 
modulus of $\Tr_m\hat\rho_{AB}(t)$, they display the {\it imaginary} part of the same quantity. 
Evaluating, like before, the decoherence-free case, we obtain
\begin{eqnarray}\label{new}
\left.\begin{array}{c}I_C\\I_D\end{array}\right\}
&=&\frac12 \left(1\,\mp\,e^{-|\alpha(t)|^2/2}\,\sin \left[\varphi(t)+\chi(t)\right]\right)\\
                 &=&\frac12 \left(1\,\mp\,e^{-\kappa^2(1-\cos \omega_m t)}\,
                      \sin \left[ \kappa^2(\omega_m t - \sin \omega_m t)+\chi(t)\right]\right),
\end{eqnarray}
where we have used Equation (\ref{phases}), and for the discussion to follow, included an additional 
phase shift $\chi(t)$ created by a phase shifter device on arm $A$ of the Michelson interferometer,
 not present in the original setup \cite{marshall}.

Having to measure the imaginary part of $\Tr_m\hat\rho_{AB}(t)$ makes a huge difference. Without 
the phase shifter, that factor oscillates as $ ~\sin \kappa^2(\omega_m t - \sin \omega_m t)$, 
which - depending on the actual value of the optomechanical coupling constant $\kappa$ - may go 
down close to zero just about time $(2\pi/\omega_m)n$, when the $n$th visibility revival is expected. 
There is a simple amendment though: one should include a constant phase shifter and tune it to 
\begin{equation}\label{first}
\chi_n=2\pi\left(\frac{2n+1}{4}-\kappa^2\right),
\end{equation}
to match the $n$th maximum of the oscillatory factor with the expected revival time. With that, revivals 
become observable in the interferometric intensity patterns, as soon as decoherence effects are suppressed 
to the desired level \cite{crosscorr}; of course, to see at least $n=1$ would be the minimum requirement.  
Alternatively, if noise level of data allows, scanning with $\chi$ to locate the maximum signal can be a way 
to measure the coupling constant $\kappa$.

The above reasoning connects the Marshall {\it et al.} scheme to a simple token introduced by  Yurke and 
Stoler \cite{cat}, used in various experiments \cite{cation,catmicro} to overcome some of the difficulties 
associated with the entanglement present e.g. in Eq. (\ref{mirrors}), and create a visible interference pattern 
of the two Schr\"odinger cat components, as apparent in Eq. (\ref{mirrorcat}), observable through 
straightforward intensity measurement of a chosen component of the associated two-state quantum system. 
In the Marshall {\it et al.} scheme it is the last passage through the Michelson beamsplitter that - acting as 
a $\pi/2$ quasi-spin rotation of the two-state photon basis - offers a kind of gratis implementation of the 
Yurke and Stoler procedure; this was the original motivation for the present work.

Alas, just seing the visibility revival would not prove quantumness \cite{bdg}. Therefore to reach that final 
scope it may be necessary to add a time-dependent phase shifter $\chi(t)$ to the photon subsystem. That 
can strongly enhance the possibilities of preparing and analysing quantum states of an oscillator, as 
demonstrated by the successful application of phase controlling protocols in ion-trap Schr\"odinger cat 
experiments \cite{cation}. Optical phase shifters based on the Pockels effect are commercially available, 
and admit electric modulation at frequencies up to a few GHz \cite{salehteich}, with the potentiality to 
make the Marshall {\it et al.} nanomechanical Schr\"odinger cat mew more clearly about his/her quantum 
behaviour. Some of the possibilities offered by that token will be analyzed in a forthcoming paper.

\ack Support by the Hungarian Scientific Research Fund OTKA under Grant No. T75129 is acknowledged. 
I thank Lajos Di\'osi for helpful discussions.


\begin{thebibliography}{9}
\bibitem{crosscorr} Another possibility is to measure cross-correlations between the two output
ports, the oscillations of which are shifted in phase by $\pi/4$ with respect to the one-port intensities.
\end{thebibliography}
\end{document}